\documentclass[twocolumn]{aastex631}

\begin{document}

\title{Asymmetric distribution of Fe-peak elements in Cassiopeia A revealed by XRISM}

\author[0000-0001-9267-1693]{Toshiki Sato}
\affil{Department of Physics, School of Science and Technology, Meiji University, Kanagawa, 214-8571, Japan}

\author[0000-0002-7273-2740]{Shin-ichiro Fujimoto}
\affil{National Institute of Technology Kumamoto College, 2659-2 Suya, Koshi, Kumamoto 861-1102, Japan}

\author[0000-0002-0018-0369]{Koji Mori}
\affil{Faculty of Engineering, University of Miyazaki, Miyazaki 889-2192, Japan}

\author{Jun Kurashima}
\affil{Faculty of Engineering, University of Miyazaki, Miyazaki 889-2192, Japan}

\author[0000-0002-0018-0369]{Hiroshi Nakajima}
\affil{College of Science and Engineering, Kanto Gakuin University, Kanagawa 236-8501, Japan}

\author[0000-0003-1415-5823]{Paul P. Plucinsky}
\affil{Center for Astrophysics, Harvard-Smithsonian, MA 02138, USA}

\author[0000-0001-6965-8642]{Manan Agarwal}
\affil{Anton Pannekoek Institute/GRAPPA, University of Amsterdam, Science Park 904, 1098 XH Amsterdam, The Netherlands}

\author[0000-0001-9911-7038]{Liyi Gu}
\affil{SRON Netherlands Institute for Space Research, Niels Bohrweg 4, 2333 CA Leiden, The Netherlands}

\author[0000-0003-3462-8886]{Adam Foster}
\affil{Center for Astrophysics, Harvard-Smithsonian, MA 02138, USA}

\author[0009-0003-0653-2913]{Kai Matsunaga}
\affil{Department of Physics, Kyoto University, Kyoto 606-8502, Japan}

\author[0000-0003-1518-2188]{Hiroyuki Uchida}
\affil{Department of Physics, Kyoto University, Kyoto 606-8502, Japan}

\author[0000-0003-0890-4920]{Aya Bamba}
\affil{Department of Physics, Graduate School of Science, The University of Tokyo, 7-3-1 Hongo, Bunkyo-ku, Tokyo 113-0033, Japan}
\affil{Research Center for the Early Universe, School of Science, The University of Tokyo, 7-3-1 Hongo, Bunkyo-ku, Tokyo 113-0033,
Japan}
\affil{Trans-Scale Quantum Science Institute, The University of T okyo, Tokyo 113-0033, Japan}

\author[0000-0002-4708-4219]{Jacco Vink}
\affil{Anton Pannekoek Institute/GRAPPA, University of Amsterdam, Science Park 904, 1098 XH Amsterdam, The Netherlands}
\affil{SRON Netherlands Institute for Space Research, Niels Bohrweg 4, 2333 CA Leiden, The Netherlands}

\author[0000-0002-2359-1857]{Yukikatsu Terada}
\affil{Graduate School of Science and Engineering, Saitama University, 255 Shimo-Ohkubo, Sakura, Saitama 338-8570, Japan}
\affil{ISAS/JAXA, 3-1-1 Yoshinodai, Chuo-ku, Sagamihara, Kanagawa 252-5210, Japan}

\author{Hironori Matsumoto}
\affil{Department of Earth and Space Science, Osaka University, Osaka 560-0043, Japan}

\author[0000-0002-5466-3817]{Lia Corrales}
\affil{Department of Astronomy, University of Michigan, MI 48109, USA}

\author[0000-0002-3844-5326]{Hiroshi Murakami}
\affil{Department of Data Science, Tohoku Gakuin University, Miyagi 984-8588, Japan}

\author[0000-0002-1104-7205]{Satoru Katsuda}
\affil{Graduate School of Science and Engineering, Saitama University, 255 Shimo-Ohkubo, Sakura, Saitama 338-8570, Japan}

\author[0000-0003-2008-6887]{Makoto Sawada}
\affil{Department of Physics, Rikkyo University, Tokyo 171-8501, Japan}

\author[0009-0006-0015-4132]{Haruto Sonoda}
\affil{Department of Physics, Graduate School of Science, The University of Tokyo, 7-3-1 Hongo, Bunkyo-ku, Tokyo 113-0033, Japan}
\affil{ISAS/JAXA, 3-1-1 Yoshinodai, Chuo-ku, Sagamihara, Kanagawa 252-5210, Japan}

\author[0000-0001-9735-4873]{Ehud Behar}
\affil{Department of Physics, Technion, Technion City, Haifa 3200003, Israel}

\author[0000-0001-7713-5016]{Masahiro Ichihashi}
\affil{Department of Physics, Graduate School of Science, The University of Tokyo, 7-3-1 Hongo, Bunkyo-ku, Tokyo 113-0033, Japan}

\author[0000-0002-5092-6085]{Hiroya Yamaguchi}
\affil{ISAS/JAXA, 3-1-1 Yoshinodai, Chuo-ku, Sagamihara, Kanagawa 252-5210, Japan}



\begin{abstract}

The elemental abundances of the Fe-peak elements (such as Cr, Mn, Fe and Ni) and Ti are important for understanding the environment of explosive nuclear burning for the core-collapse supernovae (CC SNe). In particular, the supernova remnant Cassiopeia A, which is well known for its asymmetric structure, contains three ``Fe-rich blobs,'' and the composition of the Fe-peak elements within these structures could be related to the asymmetry of the supernova explosion. We report a highly asymmetric distribution of the Fe-peak elements in Cassiopeia A as revealed by XRISM observations. We found that the southeastern Fe-rich region has a significant Mn emission above the 4$\sigma$ confidence level, while the northwestern Fe-rich region has no clear signature. In addition to the significant difference in Mn abundance across these regions, our observations show that the Ti/Fe, Mn/Cr, and Ni/Fe ratios vary from region to region. The observed asymmetric distribution of Fe-peak elements could be produced by (1) the mixing of materials from different burning layers of the supernova, (2) the asymmetric distribution of the electron fraction in the progenitor star and/or (3) the local dependence of the neutrino irradiation in the supernova innermost region. Future spatially resolved spectroscopy of Cassiopeia A using X-ray microcalorimeters will enable more detailed measurements of the distribution and composition of these elements, providing a unique tool for testing asymmetric supernova physics.

\end{abstract}



\section{Introduction} \label{sec:intro}
Core‐collapse supernova (CCSN) explosion mechanisms remain one of the great unsolved problems in astrophysics, with asymmetric dynamics thought to play a central role in driving the explosion \citep[e.g.,][]{2012ARNPS..62..407J,2021Natur.589...29B}. The $\sim$340-year-old remnant Cassiopeia A (Cas A)  provides an ideal laboratory for studying these asymmetries \citep[e.g.,][]{2000ApJ...528L.109H,2006ApJ...636..859F,2014Natur.506..339G,2015Sci...347..526M,2021Natur.592..537S,2024ApJ...965L..27M,2025A&A...696A.108O}, exhibiting three prominent Fe‐rich plumes that reveal the inhomogeneous nature of the ejecta \citep[e.g.,][]{2017ApJ...842...13W}.

The mass ratios among Fe‐peak elements are known to be key observables that characterize the silicon‐burning process ($\alpha$-rich freezeout) occurring in the innermost regions of core‐collapse supernova explosions.  In particular, during $\alpha$-rich freezeout the resulting elemental composition is governed by the electron fraction, $Y_{e}$, the peak density, the peak temperature, the rates of temperature and density evolution, and the freezeout timescale \citep[e.g.,][]{1973ApJS...26..231W,1996ApJ...460..408T}.  Measurements of these mass ratios can therefore be used to infer the neutron excess and entropy at the explosion center \citep[e.g.,][]{2015ApJ...807..110J,2021Natur.592..537S}. Moreover, recent multidimensional supernova simulations predict asymmetric distributions of $Y_{e}$ and entropy within the ejecta \citep[e.g.,][]{2018ApJ...852...40W,2018MNRAS.477.3091V,2021MNRAS.502.2319F,2024ApJ...964L..16B}, making Cas A an ideal target for testing these predicted asymmetries.

Several measurements of Fe-peak elements have been carried out in Cas A, including, for example, studies of Mn and Cr. In \cite{2020ApJ...893...49S}, the Mn/Cr ratio was used to estimate the progenitor’s initial metallicity. Here, the elements processed by the CNO cycle during stellar evolution pile up into $^{14}$N, which is burnt to $^{22}$Ne in the He burning stage through the reactions $^{14}$N($\alpha$,$\gamma$)$^{18}$F($\beta^+$)$^{18}$O($\alpha$,$\gamma$)$^{22}$Ne. The $\beta^+$ decay in this process increases the neutron excess \citep{2003ApJ...590L..83T}. Such neutronization affected by the initial metallicity also occurs in the stellar evolution of massive stars \citep[e.g.,][]{1985ApJ...295..604T}. Thus, the ratio of the neutron-rich element Mn to Cr serves as an indicator of the progenitor’s initial metallicity \cite[see also][]{2008ApJ...680L..33B,2013ApJ...767L..10P}. Based on the X-ray measurement of these elements, the authors have suggested that Cas A's progenitor formed in a sub-solar-metallicity environment. However, convective mixing within the progenitor can produce an asymmetric $Y_{e}$ distribution \citep[e.g.,][]{1994ApJ...433L..41B}, and contamination from the complete Si‐burning region near the explosion center could complicate these interpretations. Therefore, the spatial variation of the Mn/Cr ratio offers important observational insights.

\begin{figure*}[t]
 \begin{center}
  \includegraphics[bb=0 0 3133 1421,width=16cm]{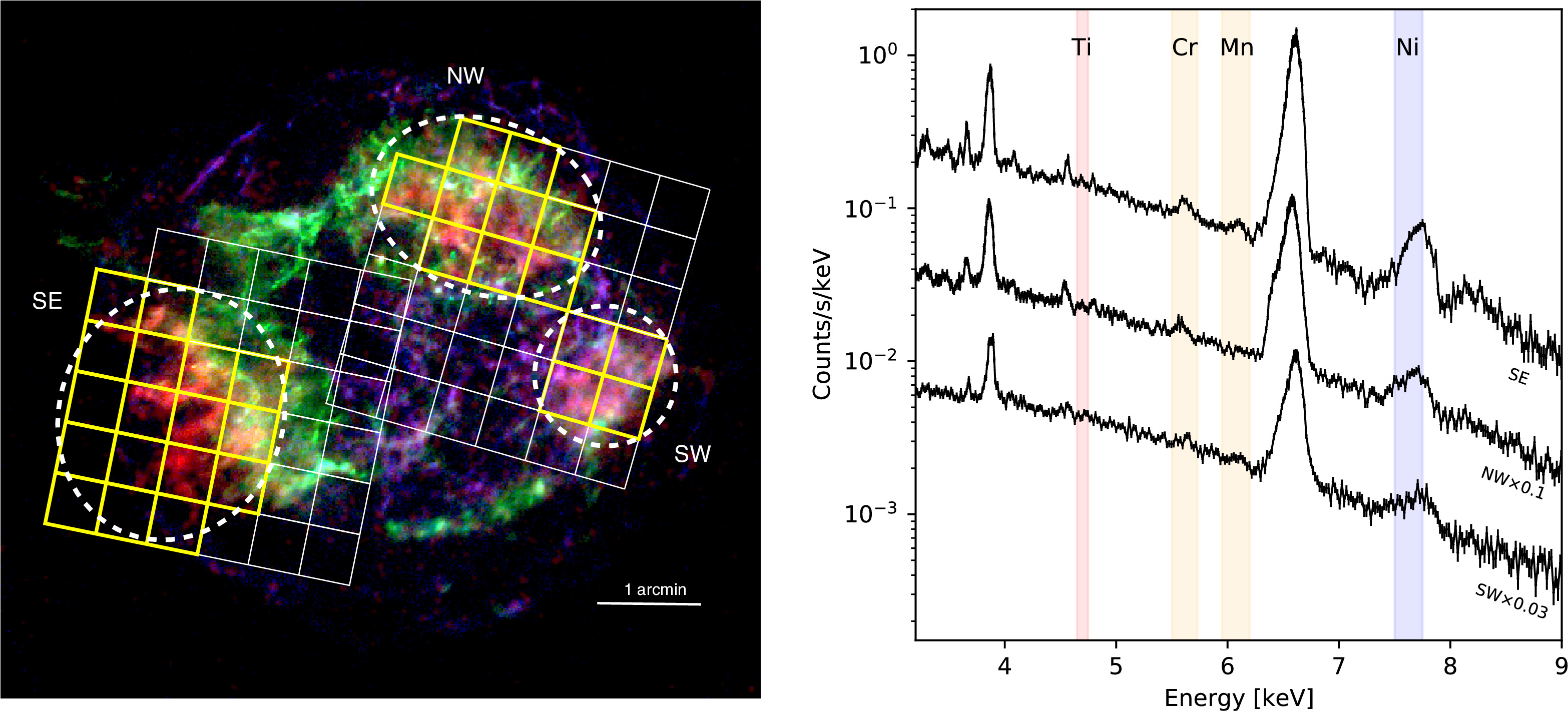}
 \end{center}
\caption{{\it Left}: The field of view of the XRISM/Resolve pointings shown on the three-color image of Cas A taken by Chandra. Red, green, and blue include emission within energy bands of 6.54--6.92 keV (Fe He$\alpha$), 1.76-1.94 keV (Si He$\alpha$), and 4.2-6.0 keV (continuum), respectively. The dashed ellipses indicate the approximate locations of the three Fe-rich blobs in Cas A.} We extracted the spectra from the pixel regions highlighted in yellow. {\it Right}: The X-ray spectra observed with XRISM/Resolve, extracted from the regions defined in the left panel. Energy bands corresponding to the He$\alpha$ emissions of Ti, Cr, Mn, and Ni are highlighted in color.
\label{fig:f1}
\end{figure*}

The synthesized yield of Mn is also linked to neutrino irradiation near the explosion center, and \cite{2023ApJ...954..112S} have discussed this possibility within Cas A's Fe-rich ejecta.
Neutrinos can excite heavy nuclei, which subsequently emit nucleons and light nuclei, significantly altering the final composition. This process is called the ``$\nu$-process'' \citep[e.g.,][]{1988Natur.334...45W,1990ApJ...356..272W,1995ApJS..101..181W,2005PhLB..606..258H,2005PhRvL..94w1101Y,2008ApJ...672.1043Y,2013ApJ...779L...9H,2018PhRvL.121j2701H,2018ApJ...865..143S,2019ApJ...876..151S}. In the $\nu$-process, neutrino spallation, working on the $^{56}$Ni produced by complete and incomplete Si burning, enhances the Mn abundance through the reaction $^{56}{\rm Ni}(\nu,\nu^\prime p)^{55}{\rm Co}$, where $^{55}$Co decays to $^{55}$Fe and then to stable $^{55}$Mn. Especially in the complete Si burning regime, the production of Mn without the $\nu$-process is much smaller than with it \citep[as we show below, and see also][]{2008ApJ...672.1043Y}. In addition to the $\nu$-process, the strongly neutrino-processed proton-rich environment can also enhance the Mn abundance at the complete Si burning region \citep{2018ApJ...852...40W}, where $^{55}$Mn is initially produced as the proton-rich isotope $^{55}$Ni, which then decays via Co and Fe to $^{55}$Mn. In this case, the innermost material is strongly irradiated by neutrinos during the accretion phase, producing a proton-rich environment that synthesizes $^{55}$Ni effectively. After the accretion phase and the SN shock passage, neutrinos still irradiate the central materials during the cooling phase, so the $^{56}{\rm Ni}(\nu,\nu^\prime p)^{55}{\rm Co}$ reaction increases the amount of Mn further. Thus, both the $\nu$-process and a proton-rich environment (i.e., the $\nu$p-process) can enhance Mn production, which means that we can test the existence of the neutrino interaction by measuring the Mn abundance in the innermost ejecta. 

Not only Mn and Cr, but also Ti and Ni, which are also synthesized at the explosion center, have attracted attention as tracers of entropy and neutron excess. \cite{2021Natur.592..537S} have detected Ti in the Fe-rich structure in Cas A’s southeast region and, based on its abundance, showed that this feature represents a high-entropy ejecta plume produced by the neutrino-heating mechanism. Although Ni has similarly been identified, accurately quantifying its synthesized mass has been challenging with previous detectors \citep[see][]{2021Natur.592..537S}. With the high-resolution X-ray spectroscopy of XRISM’s Resolve microcalorimeter \citep{2025PASJ..tmp...28T,ishisaki2025}, we can now measure these elements from Ti through Ni with unprecedented precision and probe their spatial variations. In this paper, we investigate the asymmetric distribution of Fe-peak elements in Cas A based on early XRISM satellite observations.

\begin{figure*}[t]
 \begin{center}
  \includegraphics[bb=0 0 2792 1196,width=18cm]{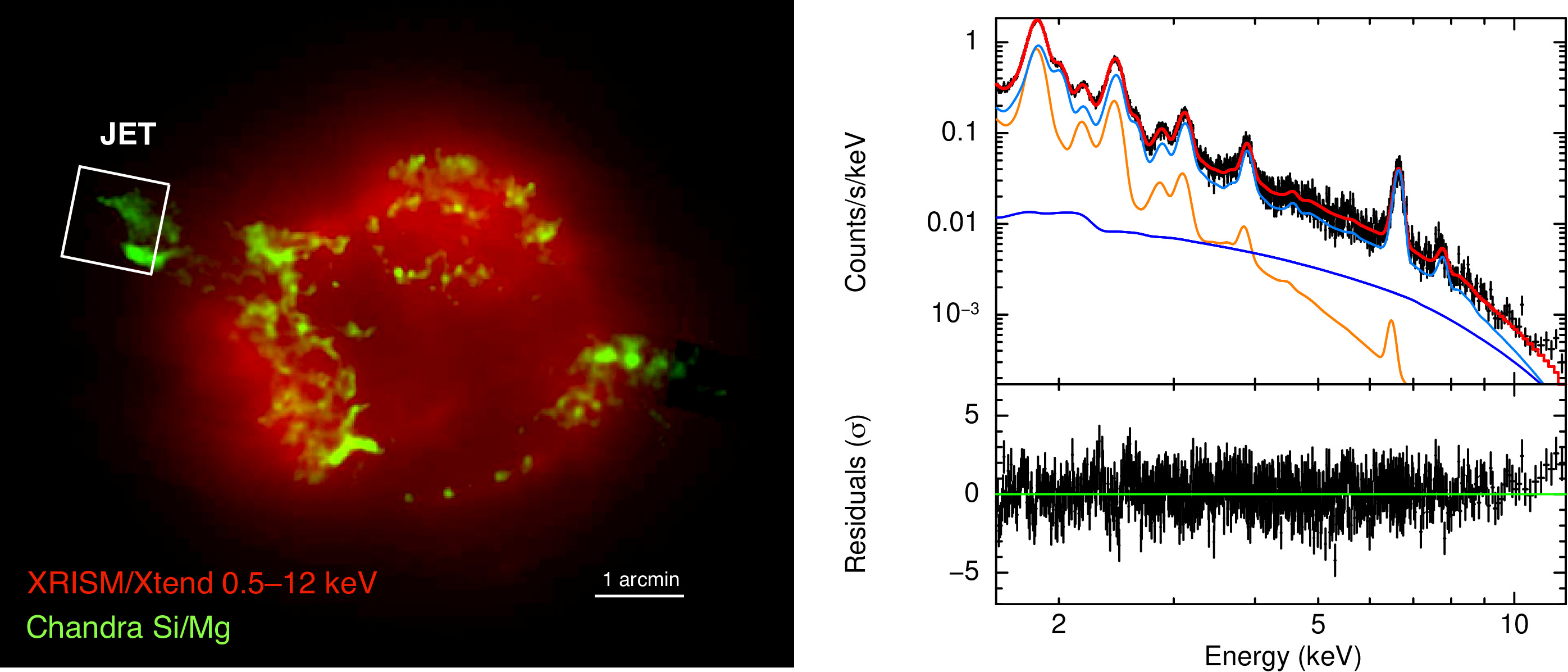}
 \end{center}
\caption{{\it Left}: Two-color image of Cas A obtained with XRISM/Xtend and Chandra. Red indicates the 0.5–12 keV emission observed with Xtend, and green shows the Si-to-Mg line emission ratio map. The spectra were extracted from the white-box region labeled “JET.”
{\it Right}: X-ray spectrum from the JET region (black points) along with the best-fit model (red solid line) in the 1.6–12 keV energy range. The lower panel displays the residuals between the data and the model. The orange, light blue, and blue solid lines correspond to the low-temperature plasma component, the high-temperature plasma component, and the power-law component, respectively.}
\label{fig:f2-2}
\end{figure*}

\begin{figure*}[t]
 \begin{center}
  \includegraphics[bb=0 0 3371 1521, width=18cm]{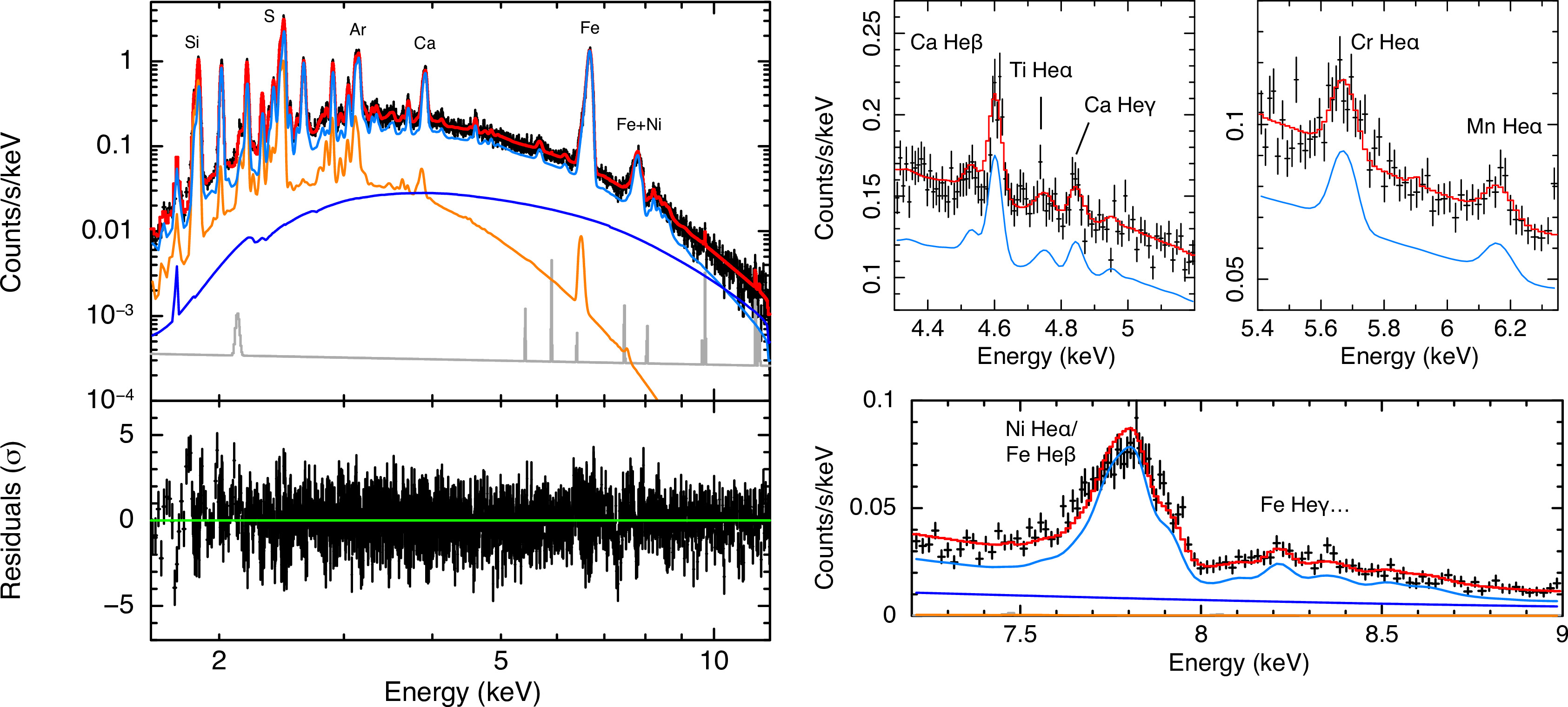}
 \end{center}
\caption{X-ray spectrum of the SE region obtained with Resolve (black points) together with the best-fit model (red solid line). The left panel presents the fit in the 1.6–12 keV energy range, while the lower panel shows the residuals between the data and the model. The orange, light blue, blue, and gray solid lines represent the low-temperature plasma model, the high-temperature plasma model, the power-law model, and the NXB model, respectively. The multiple panels on the right show zoomed-in views of the energy bands of Ti, Cr, Mn, and Ni.}
\label{fig:f2}
\end{figure*}

\begin{table*}[t]
    \centering
    \caption{Best-fit paremeters of spectral fits.}
    \begin{tabular}{lccccccc}
    \hline
    Region & \multicolumn{2}{c}{SE} & \multicolumn{2}{c}{NW} & \multicolumn{2}{c}{SW} & \multicolumn{1}{c}{JET}\\
       Atomic Database & AtomDB & SPEX & AtomDB & SPEX & AtomDB & SPEX & AtomDB\\
    \hline
Fitting parameters\\
(Mg/H)/(Mg/H)$_\odot$ & 3.8$^{+0.4}_{-0.2}$     & 5.3$^{+0.7}_{-0.6}$   & 1.8$^{+0.3}_{-0.2}$    & 3.6$^{+0.6}_{-0.5}$    & 1.3$\pm$0.3            & 3$\pm$1            & 6.5$^{+0.5}_{-0.6}$           \\
(Al/H)/(Al/H)$_\odot$ & 5$\pm$1                 & 3$\pm$1               & 3.4$^{+0.8}_{-0.7}$    & 3$\pm$1                & $<$0.5                 & $<$2               &    ---      \\
(Si/H)/(Si/H)$_\odot$ & 7.5$^{+0.2}_{-0.3}$     & 7.5$^{+0.4}_{-0.3}$   & 4.32$^{+0.09}_{-0.06}$ & 5.3$\pm$0.2            & 1.91$^{+0.06}_{-0.09}$ & 2.3$\pm$0.2        & 8.0$^{+0.7}_{-0.4}$     \\
(P/H)/(P/H)$_\odot$   & 10$\pm$1                & 6$\pm$1               & 4.7$\pm$0.8            & 3$\pm$1                & 4$\pm$1                & 5$\pm$2            &   ---  \\
(S/H)/(S/H)$_\odot$   & 7.8$^{+0.2}_{-0.1}$     & 7.2$\pm$0.3           & 5.20$^{+0.10}_{-0.08}$ & 5.8$\pm$0.2            & 2.31$^{+0.06}_{-0.09}$ & 2.5$\pm$0.2        & 8.0$\pm$0.6\\
(Cl/H)/(Cl/H)$_\odot$ & 11.8$\pm$0.9            & 7.4$^{+1.0}_{-0.9}$   & 7.4$^{+0.7}_{-0.6}$    & 6.1$\pm$0.8            & 3.2$\pm$0.7            & 3$\pm$2            &   ---         \\
(Ar/H)/(Ar/H)$_\odot$ & 7.0$\pm$0.2             & 6.0$\pm$0.3           & 4.81$^{+0.11}_{-0.06}$ & 5.3$\pm$0.1            & 2.36$^{+0.07}_{-0.09}$ & 2.6$\pm$0.2        & 7.5$\pm$0.6 \\
(K/H)/(K/H)$_\odot$   & 9$\pm$1                 & 8$\pm$1               & 5.3$\pm$0.7            & 6.1$\pm$0.9            & 1.8$\pm$0.8            & 2$\pm$2            &  --- \\
(Ca/H)/(Ca/H)$_\odot$ & 7.0$^{+0.4}_{-0.2}$     & 6.3$\pm$0.3           & 5.27$^{+0.13}_{-0.07}$ & 5.8$\pm$0.2            & 2.75$^{+0.08}_{-0.12}$ & 3.0$^{+0.3}_{-0.2}$    & 9.0$\pm$0.8                \\
(Ti/H)/(Ti/H)$_\odot$ & 4$\pm$2                 & 6$\pm$2               & $<$0.5                 & $<$0.3                 & 3$\pm$2                & $<$4               & $<$1.7          \\
(Cr/H)/(Cr/H)$_\odot$ & 5.0$\pm$0.6             & 4.9$^{+0.7}_{-0.6}$   & 3.1$\pm$0.5            & 3.2$\pm$0.6            & 3.0$\pm$0.7            & 3$^{+2}_{-1}$      & $<$4.3            \\
(Mn/H)/(Mn/H)$_\odot$ & 4$\pm$1                 & 4$\pm$1               & $<$0.3                 & $<$0.5                 & 2$\pm$1                & $<$6               &   ---                    \\
(Fe/H)/(Fe/H)$_\odot$ & 11.4$^{+0.2}_{-0.4}$    & 7.9$^{+0.6}_{-0.5}$   & 6.64$^{+0.08}_{-0.05}$ & 6.2$\pm$0.3            & 3.5$\pm$0.2            & 3.0$^{+0.4}_{-0.3}$    &3.0$\pm$0.4        \\
(Ni/H)/(Ni/H)$_\odot$ & 15$\pm$1                & 5.3$^{+0.7}_{-0.6}$   & 7.5$^{+0.7}_{-0.8}$    & 3.5$\pm$0.6            & 6$\pm$1                & 3$^{+2}_{-1}$      &7.1$^{+2.7}_{-2.6}$         \\
$kT_{e,1}$ [keV]      & 0.89$^{+0.01}_{-0.02}$  & 0.75$\pm$0.2          & 0.69$^{+0.02}_{-0.01}$ & 0.58$^{+0.03}_{-0.02}$ & 0.87$^{+0.03}_{-0.06}$ & 0.82$\pm$0.06      & 1.1$\pm$0.1      \\
$kT_{e,2}$ [keV]      & 2.41$\pm$0.02           & 2.90$^{+0.09}_{-0.08}$& 2.54$^{+0.07}_{-0.03}$ & 2.37$\pm$0.05          & 2.34$^{+0.09}_{-0.07}$ & 2.3$\pm$0.2        & 3.1$^{+0.2}_{-0.3}$         \\
$n_e t_1$ [10$^{10}$ cm$^{-3}$ s] & 4.9$\pm$0.3 & 32$^{+4}_{-3}$        & 4.1$\pm$0.3            & 5.2$^{+0.6}_{-0.5}$    & 60$^{+50}_{-10}$       & $>$50              & 5.6$^{+4.9}_{-2.5}$               \\
$n_e t_2$ [10$^{11}$ cm$^{-3}$ s]&2.30$^{+0.05}_{-0.03}$& 2.02$\pm$0.06 & 1.33$^{+0.02}_{-0.01}$ & 1.55$\pm$0.04          & 1.83$\pm$0.08          & 2.1$^{+0.3}_{-0.2}$    & 1.7$^{+0.2}_{-0.1}$             \\
Redshift $z_1$ [10$^{-3}$]     & $-$1.6$\pm$0.1 & $-$1.6$\pm$0.1        & 0.07$\pm$0.01          & 6.09$\pm$0.07          & 0.51$^{+0.07}_{-0.08}$ & 0.5$\pm$0.1        & 4.3$^{+1.2}_{-0.9}$             \\
Redshift $z_2$ [10$^{-3}$]   & $-$4.15$\pm$0.04 & $-$4.46$\pm$0.06      & 6.36$\pm$0.08          & $-$0.1$\pm$0.3         & 4.2$\pm$0.1            & 4.2$\pm$0.3        & $-$1.9$\pm$0.4       \\
RMS $V_1$ [km s$^{-1}$]      & 1150$\pm$30      & 1230$\pm$30           & 1720$\pm$50            & 1730$^{+70}_{-60}$     & 490$^{+20}_{-30}$      & 500$\pm$50         &  ---       \\
RMS $V_2$ [km s$^{-1}$]      & 1290$\pm$10      & 1250$\pm$10           & 1830$\pm$10            & 1840$\pm$10            & 1850$\pm$40            & 1890$\pm$70        &   ---       \\
$\frac{n_{\rm e}n_{\rm p}V_{\rm NEI,1}}{4\pi D^2}$ [$10^{-14}$~cm$^{-5}$]&0.70$^{+0.08}_{-0.02}$ & 1.4$\pm$0.1 &   1.88$^{+0.07}_{-0.05}$   & 2.6$\pm$0.3               &   1.6$^{+0.3}_{-0.1}$  & 2.0$^{+0.5}_{-0.4}$   &  0.5$^{+0.2}_{-0.1}$             \\
$\frac{n_{\rm e}n_{\rm p}V_{\rm NEI,2}}{4\pi D^2}$ [$10^{-14}$~cm$^{-5}$]&  0.50$\pm$0.01        & 0.45$\pm$0.02 &   0.86$\pm$0.01          & 0.87$\pm$0.02         &   1.41$^{+0.04}_{-0.03}$   &  1.4$\pm$0.1  &  0.50$\pm$0.02            \\
Power-law index & 2.3 & 2.3 & 2.3 & 2.3 & 2.3 & 2.3 & 2.3 \\
Power-law norm  &  0.055$^{+0.004}_{-0.005}$   &  0.022$\pm$0.007 &  0.077$^{+0.002}_{-0.001}$   &  0.088$\pm$0.003   &  0.26$^{+0.01}_{-0.02}$   & 0.26$\pm$0.03  & 0.07$\pm$0.02\\
C-value/d.o.f         & 3652.63/3001            & 3748.35/3003          & 3847.30/3065           & 4165.77/3066           & 3214.40/2999           & 3316.26/3000       & 961.19/825           \\ \hline
Mass ratios &  &  &  &  &  & \\
$M_{\rm Ca}/M_{\rm Fe}$ [10$^{-2}$] & 3.2$^{+0.2}_{-0.1}$    & 4.1$^{+0.4}_{-0.3}$ & 4.0$\pm$0.1    & 4.8$^{+0.3}_{-0.2}$   & 4.0$^{+0.2}_{-0.3}$ & 5.1$\pm$0.1 & 15$\pm$2\\
$M_{\rm Ti}/M_{\rm Fe}$ [10$^{-3}$] & 0.9$\pm$0.4            & 1.8$\pm$0.6         & $<$1.9         & $<$1.4                & 2.2$\pm$1.4 & $<$5.9  & $<$1.4\\
$M_{\rm Cr}/M_{\rm Fe}$ [10$^{-3}$] & 4.8$\pm$0.8            & 6.8$\pm$1.3         & 5.1$\pm$1.1    & 5.8$\pm$1.4           & 9.6$\pm$3.0 & 10.8$^{+8.0}_{-6.3}$ & $<$21\\
$M_{\rm Mn}/M_{\rm Cr}$ [10$^{-1}$] & 6.5$\pm$1.7            & 6.8$\pm$2.0         & $<$0.8         & $<$1.1                & 4.9$\pm$3.1 & $<$15 & ---\\
$M_{\rm Ni}/M_{\rm Fe}$ [10$^{-2}$] & 8.1$^{+0.6}_{-0.7}$    & 4.1$\pm$0.6         & 6.7$\pm$0.7    & 6.8$\pm$2.0           & 11$\pm$2 & 5.5$^{+3.3}_{-2.4}$ & 14$\pm$6\\
    \hline
    \end{tabular}
    \label{tab:best-fit}
    
    \noindent\raggedright \textbf{Note.} The error shows 1$\sigma$ confidence level. The proto-solar abundances in \cite{2009LanB...4B..712L} are used. The abundances are linked between the low-temperature (1) and high-temperature (2) components. For the analysis of the JET region with Xtend, the data were binned to ensure at least 20 counts per energy bin, and the spectra were fitted using $\chi^2$ statistics. Even using the \textit{C}-statistics, the values are consistent within the statistical error ranges. The unit of the power-law norm is [photons/keV/cm$^2$/s] at 1 keV. Even when using a softer power-law index of 2.8 instead of the fixed power-law index (see Table \ref{tab:massratio}), the conclusions of this study remain unchanged.
\end{table*}

\begin{table*}[t]
\caption{Mass ratios with different fitting conditions.}
\begin{tabular}{c c c c c c c}
\hline
Region ID.
&
& $M_{\rm Ca}/M_{\rm Fe}\,[10^{-2}]$
& $M_{\rm Ti}/M_{\rm Fe}\,[10^{-3}]$
& $M_{\rm Cr}/M_{\rm Fe}\,[10^{-3}]$
& $M_{\rm Mn}/M_{\rm Cr}\,[10^{-1}]$
& $M_{\rm Ni}/M_{\rm Fe}\,[10^{-2}]$ \\
\hline
\multicolumn{4}{l}{Soft power-law index/AtomDB fit}\\
\multicolumn{2}{c}{SE} & $3.1^{+0.2}_{-0.4}$ & $0.9\pm0.4$ & $6.1^{+0.8}_{-1.1}$ & $6.2^{+1.8}_{-1.7}$ & $8.2^{+2.0}_{-1.3}$ \\
\multicolumn{2}{c}{NW} & $4.1\pm0.1$ & $<0.2$ & $6.8\pm1.2$ & $<1.0$ & $6.5\pm0.7$ \\
\multicolumn{2}{c}{SW} & $4.2\pm0.6$ & $2.5^{+1.4}_{-1.5}$ & $12\pm3$ & $4.1\pm3.1$ & $10\pm4$ \\
\multicolumn{4}{l}{Soft power-law index/SPEX fit}\\
\multicolumn{2}{c}{SE} & $4.1\pm0.7$ & $1.8\pm0.6$ & $8.9^{+2.3}_{-2.2}$ & $6.9^{+2.7}_{-2.5}$ & $4.1^{+1.1}_{-1.0}$ \\
\multicolumn{2}{c}{NW} & $5.0\pm0.4$ & $<0.2$ & $8.4\pm1.4$ & $<1.4$ & $3.0\pm0.6$ \\
\multicolumn{2}{c}{SW} & $5.3\pm0.9$ & $2.3^{+2.0}_{-2.1}$ & $14^{+4}_{-5}$ & $5.3^{+4.8}_{-4.6}$ & $5.2\pm1.9$ \\
\multicolumn{3}{l}{$N_{\rm H}$ free/AtomDB fit}\\
\multicolumn{2}{c}{SE} & $3.0\pm0.3$ & $1.1\pm0.5$ & $6.5^{+0.6}_{-0.8}$ & $6.7\pm1.7$ & $7.9^{+0.7}_{-1.0}$ \\
\multicolumn{2}{c}{NW} & $4.5\pm0.2$ & $<0.5$ & $8.1^{+1.4}_{-1.3}$ & $<1.2$ & $4.6\pm0.6$ \\
\multicolumn{2}{c}{SW} & $4.1\pm0.3$ & $2.4\pm1.6$ & $12\pm4$ & $4.5^{+3.2}_{-3.1}$ & $10\pm2$ \\
\hline
\end{tabular}
\label{tab:massratio}

    \vspace{0.1cm}
    \noindent\raggedright \textbf{Note.} The error shows 1$\sigma$ confidence level. In the ``soft power-law index'' fit, the power-law index was fixed at 2.8.
\end{table*}

\section{Observation and Data Reduction} \label{sec:obs}
We performed a two‐pointing observation of Cas A during the XRISM commissioning phase. From these observations, we have already reported on the impact of atomic data and spectral models on derived abundances \citep{2025PASJ...77S.171P}, as well as on the measurement of the remnant's expansion \citep{2025PASJ..tmp...58B,2025PASJ..tmp...60S,2025PASJ...77S.154V} and the odd-Z elements \citep{2025NatAs.tmp..243X} enabled by the Resolve's high-energy resolution. While \cite{2025PASJ...77S.171P} also reported spatial variations in the abundance ratios of the Fe-peak elements, in this paper we present a more detailed analysis and interpretation, with a particular focus on the abundance ratios in the Fe-rich structures. The Resolve fields of view (FoVs) are shown in Fig.~\ref{fig:f1} (left). Data reduction utilized calibration files from the HEASARC Calibration Database (CALDB). Cleaned event lists were produced with HEASARC software v6.34, applying standard screening during post‐pipeline processing, resulting in clean exposure times of 181.3 and 165.7 ks for SE and NW, respectively. For spectral analysis, we selected only the highest‐resolution (``Hp'') primary events. The redistribution matrix file (RMF) was generated in “extra‐large” mode via \texttt{rslmkrmf}, and the ancillary response file (ARF) was generated with \texttt{xaarfgen}, adopting Cas A’s surface brightness profile as measured from a 2.0--8.0 keV Chandra X-ray image.

We defined the three Fe-rich regions observed in Cas A (southeast: SW, northwest: NW and southwest: SW) as pixel regions in XRISM/Resolve and extracted the spectrum from each (Fig.~\ref{fig:f1}, right). The high-energy-resolution spectra obtained with Resolve clearly show that the intensity ratios of Cr, Mn, and Ni emission lines vary between regions. In particular, Mn and Ni lines are enhanced in the SE region and suppressed in the NW region, indicating an asymmetric distribution of Fe-peak elements. To verify that these line-ratio variations correspond to true abundance differences, we derived elemental abundances via spectral fitting. Each fit was performed with SPEX v3.08.01 \citep{kaastra1996} and XSPEC v12.14.1 (AtomDB v3.1.3) \citep{arnaud1996}, using the maximum-likelihood \textit{C}-statistic \citep{cash1979} over the 1.6--12.0 keV band. To account for the non-X-ray background (NXB), we applied the temporal NXB spectral model provided by the XRISM calibration team, constructed from 785 ks of stacked night-Earth observations. This model comprises a power law plus 17 Gaussian components for Al K$\alpha_{1}$/K$\alpha_{2}$, Au M$\alpha_{1}$, Cr K$\alpha_{1}$/K$\alpha_{2}$, Mn K$\alpha_{1}$/K$\alpha_{2}$, Fe K$\alpha_{1}$/K$\alpha_{2}$, Ni K$\alpha_{1}$/K$\alpha_{2}$, Cu K$\alpha_{1}$/K$\alpha_{2}$, and Au L$\alpha_{1}$/L$\alpha_{2}$, L$\beta_{1}$/L$\beta_{2}$. We note that the level of the NXB is more than order of magnitude below the source spectrum in the energy ranges we are interested in. Thus, our results are insensitive to any uncertainty in the NXB model.

We also investigated the jet structure outside the field of view of Resolve by analyzing the data obtained by XRISM/Xtend \citep{2024SPIE13093E..1IM} (Figure~\ref{fig:f2-2}). As with Resolve, we applied the standard screening, which resulted in a clean exposure time of 215.2 ks for the SE pointing,  and generated the RMF file using \texttt{xtdrmf} and the ARF file using \texttt{xaarfgen}. We generated the background spectrum of the analysis region using the NXB database, and subtracted it from the source region spectrum. In the NW pointing, a charge injection (CI) row intersects the center of the jet structure, which makes the evaluation difficult. Therefore, for Xtend, we analyze only the SE pointing.

\section{Asymmetric distribution of Fe-peak elements} \label{sec:asymmetric}

We present here a detailed evaluation of the X-ray spectra and the corresponding observational results. Figure~\ref{fig:f2} presents the results of plasma modeling of the Resolve's X-ray spectrum in the SE region, and the best-fit parameters are summarized in Table~\ref{tab:best-fit}. Based on previous observations \citep{2012ApJ...746..130H}, we fitted the spectra with a two-temperature non-equilibrium ionization (NEI) model, absorbed by the interstellar medium (ISM) with solar abundances \citep{2009LanB...4B..712L}. The high-temperature ejecta component was modeled with a plane-parallel shock (\texttt{bvvpshock} in Xspec and \texttt{neij} with `mode=3' in SPEX) plasma. Here, the fitting model in Xspec is \texttt{phabs*(bvvnei+bvvpshock+pow)} and that in SPEX is \texttt{abs*(neij(mode=1)+neij(mode=3)+pow)}. We allowed the volume emission measure ($=n_{\rm e}n_{\rm p}V_{\rm NEI}$, where $n_{\rm e}$, $n_{\rm p}$, and $V_{\rm NEI}$ are the electron density, proton density, and plasma volume, respectively), the electron temperature, $kT_{\rm e}$, and the ionization timescale, $n_{\rm e}t$, line broadening and redshift parameters to vary. Abundances of Mg, Al, Si, P, S, Cl, Ar, K, Ca, Ti, Cr, Mn, Fe, and Ni were left free and tied between both NEI components, while all other elements were fixed at solar values. As shown in Figure~\ref{fig:f2}, when the spectra are fitted with two plasma components at different temperatures, nearly all of the emission from the Fe-group elements that are the focus of this study originates from the high-temperature component. Therefore, even if the abundances are not tied between the two NEI components, our results remain unchanged. The absorption column density, $N_{\rm H}$, was fixed at $1.9\times10^{22}$, $2.0\times10^{22}$, and $1.5\times10^{22}$ cm$^{-2}$ for the SE, NW, and SW regions, respectively, with reference to the previous observational result \citep{2012ApJ...746..130H}. Even if $N_{\rm H}$ is treated as a free parameter, our conclusions remain unchanged (Table~\ref{tab:massratio}). In the case where $N_{\rm H}$ is allowed to vary freely, the best-fit values of $N_{\rm H}$ in the SE, NW, and SW regions are estimated to be $N_{\rm H} = (7.3\pm0.8)\times10^{21}$, $(3.7^{+0.7}_{-1.0})\times10^{21}$, and $(1.6^{+0.1}_{-0.2})\times10^{22}$ cm$^{-2}$, respectively, where the results differ significantly from previous studies in all regions except the SW region. During the XRISM/Resolve observations, the gate valve was closed, and photons below $\sim$1.5~keV were scarcely detected.

Based on the detailed spectral analysis, we found that the elemental composition ratios vary within the Fe-rich structures of Cas A. The clearest difference appears between the SE region and the NW region, and the variation in Mn abundance is especially noteworthy (Table~\ref{tab:best-fit}). In the SE region, Mn is detected with a confidence level greater than $4\sigma$ (see Figure~\ref{fig:f1}), whereas in the NW region Mn is not detected even at the $1\sigma$ level. This makes it clear that the Mn abundance differs between these two areas. Moreover, when we compare the ratio of Mn to Cr, which is the element with the neighboring atomic number that should be synthesized in the same region, we find a Mn/Cr mass ratio of 0.65--0.68 in the SE region and Mn/Cr $<$ 0.11 in the NW region (Table~\ref{tab:best-fit}), showing a significant difference in their relative abundances. The result for the SW region lies between those for the SE and NW regions, suggesting that the Mn abundance varies from region to region in Cas A. In particular, as a neutron‐rich odd‐Z element, Mn serves as a tracer of the electron fraction and neutrino interactions in the nuclear‐burning regions, making this asymmetric distribution especially intriguing (see section~\ref{sec:discussion} for detail). The trend in the Mn abundance variation between the SE and NW regions remains consistent across different fitting conditions (Table~\ref{tab:best-fit}), making it one of the most robust results of this study.

\begin{figure*}[t]
 \begin{center}
  \includegraphics[bb=0 0 3463 1494, width=18cm]{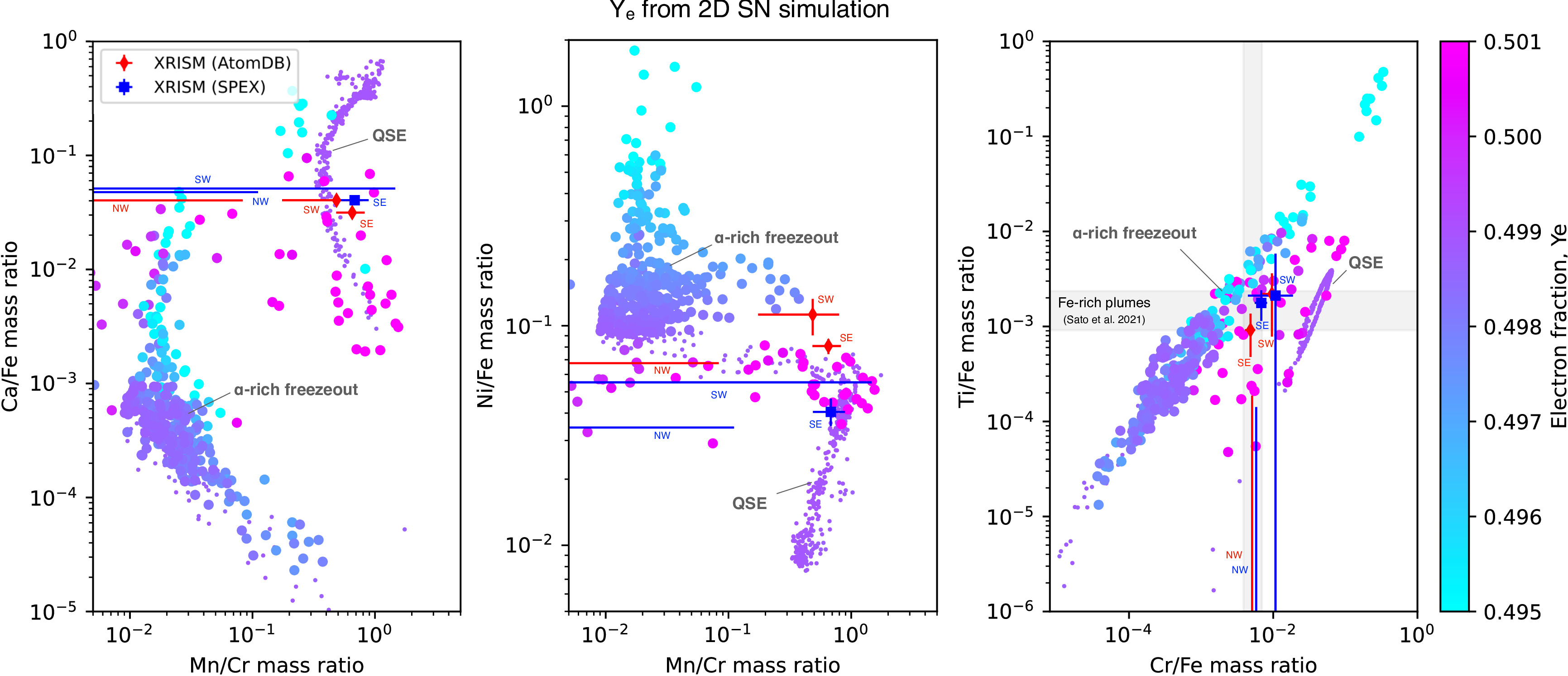}
  \par \vspace{0.2cm}
  \includegraphics[bb=0 0 3463 1494, width=18cm]{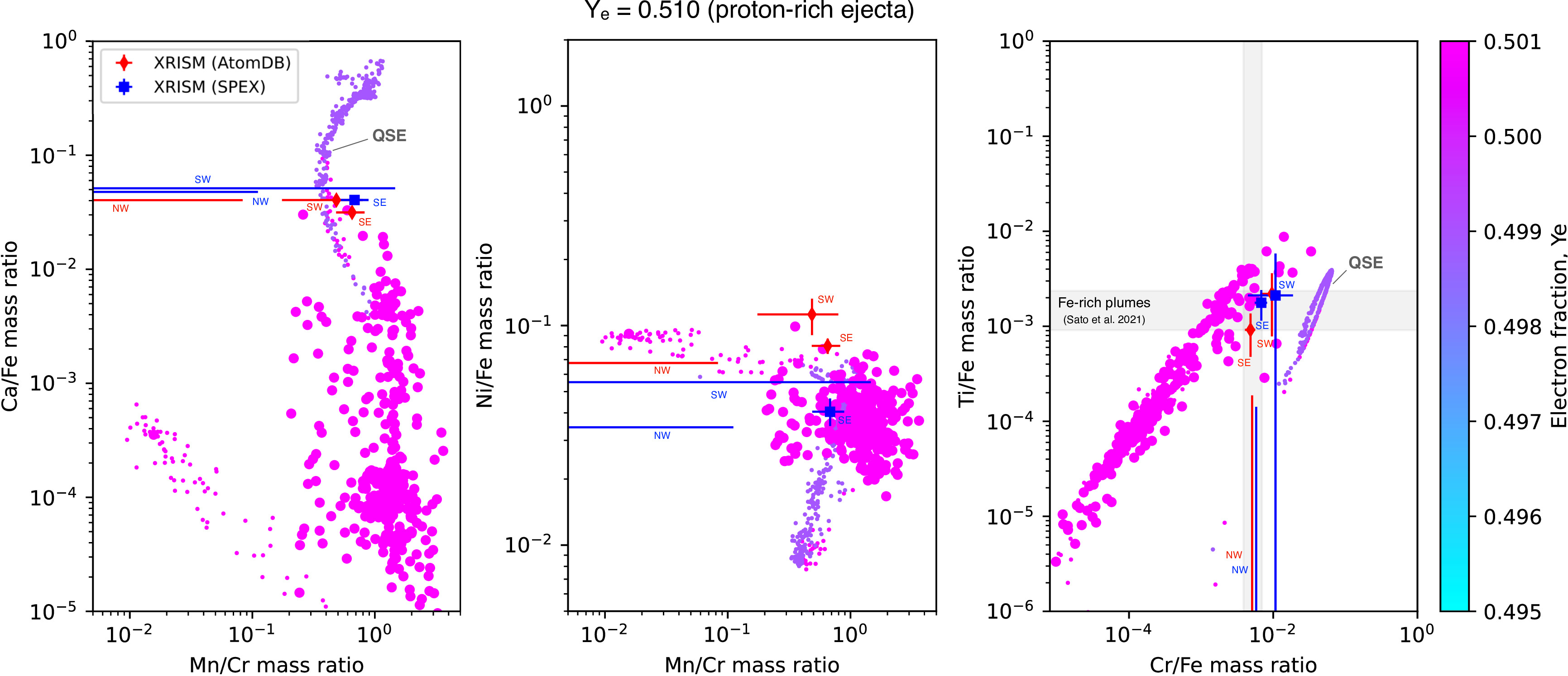}
 \end{center}
\caption{{\it Top panels}: Comparison of the elemental mass ratios in individual fluid elements of a 2D supernova model with those observed in Cas A. From left to right, the Mn/Cr versus Ca/Fe plot, the Mn/Cr versus Ni/Fe plot, and the Cr/Fe versus Ti/Fe plot are shown. Model points are colored by their electron fraction ($Y_e$), with light blue indicating low $Y_e$ regions and magenta indicating high $Y_e$ regions. Data points for $T_{\rm peak} > 5.5$ GK (i.e., complete Si burning or $\alpha$-rich freezeout) are represented by large circles, whereas those for $T_{\rm peak} < 5.5$ GK (i.e., incomplete Si burning or quasi-equilibrium: QSE) are represented by small circles. Observational values appear as red and blue data points with error bars (1$\sigma$ confidence level). In the Ti/Fe vs. Cr/Fe plot, the observational values reported in \cite{2021Natur.592..537S} are indicated by gray areas. {\it Bottom panels}: Same as the top panels, but compared with nucleosynthesis calculations with the electron fraction fixed at $Y_e = 0.51$, resembling a case where most of the ejecta become proton-rich ($Y_e > 0.5$) owing to neutrino irradiation. The electron fraction is fixed only for ejecta with peak temperatures exceeding 5.5 GK. The observational values in these plots are taken from Table~\ref{tab:best-fit}.}
\label{fig:f3}
\end{figure*}

In addition to Mn, the abundance of another key neutron-rich element, Ni, also appears to vary among the regions. However, the uncertainties in plasma modeling for Ni remain large, and at present different atomic codes yield significantly different abundance estimates \citep{2025PASJ...77S.171P}. This uncertainty arises primarily from the broad line profiles, which prevent clear separation of Fe He$\beta$ emission from Ni He$\alpha$ emission (see the zoom-up panel in Figure \ref{fig:f2}). In addition, differences between atomic codes in the emissivities of the various lines within this band further complicate these estimates. Determining the absolute Ni/Fe ratio is critical for inferring the physical conditions at the explosion center (see section \ref{sec:discussion}), yet with the current data set an accurate estimate is challenging. We hope that future detailed non-equilibrium plasma modeling will help resolve these issues.

Although the detection significance is modest, the detection of Ti in the SE region at roughly the $3\sigma$ level is noteworthy (see the zoom-up panel around the Ti He$\alpha$ emission in Figure \ref{fig:f2}). The observed Ti/Fe ratio is 0.9--1.8, which agrees with the previous measurement with Chandra \citep[see][and section \ref{sec:discussion} for the detailed discussion]{2021Natur.592..537S}. The energy scale for the Cas A observations was calibrated using on-board $^{55}$Fe sources, and the resulting systematic uncertainty in the reconstructed energy scale is $<$0.04 eV at 5.9 keV for the full detector array \citep[see Methods in][for more detail]{2025NatAs.tmp..243X}. In addition, the effect of thermal broadening near the Ti He$\alpha$ emissions is $\sim$20 eV, which is larger than the Resolve's energy resolution of $\sim$5 eV. Under the current conditions of energy-scale reconstruction and line broadening, the separation of the Ti He$\alpha$ line from the neighboring Ca He$\beta$/$\gamma$ line complex on either side is achievable, enabling a more precise measurement than was possible with the previous Chandra observations. In the NW and SW regions, Ti is not detected at a statistically significant level (at $<1\sigma$ level). One reason may be that, compared to the SE region, these regions have a larger contribution from non-thermal emission, resulting in a lower signal-to-noise ratio. In these regions, we also found that the detection significance or upper limit for Ti changes depending on variations in the parameters of the interstellar absorption and the power-law component (Table~\ref{tab:massratio}). In the future, deeper observations will be necessary to determine the Ti abundance in those regions. 

In some fittings, the temperature and ionization age of the low-temperature component differ between AtomDB and SPEX. This is likely due to the difficulty of fitting caused by the lack of emission below 1 keV. On the other hand, the emissions from the Fe-group elements mainly depends on the high-temperature component, and we have confirmed that the uncertainties in the low-temperature component's parameters do not affect our conclusions.

We also investigated the presence of Fe-peak elements in the northeastern jet structure using XRISM/Xtend. As in the other regions, the spectral fitting was performed with a model consisting of a two-temperature plasma component plus an additional power-law. For the abundances of odd-Z elements and the line broadening, which are difficult to constrain with Xtend, the parameters were fixed to the values obtained in the SE region with Resolve. Although weak emission lines such as Ti, Cr, and Mn were not significantly detected, the low background level enabled us to identify a prominent emission-line structure around the Ni band. Consequently, the analysis yields a relatively large mass ratio of Ni/Fe = 0.14 $\pm$ 0.6. While systematic uncertainties among atomic codes remain substantial for Ni, this result may suggest a contribution from the $\alpha$-rich freezeout region near the explosion center \citep[see also][]{2025ApJ...990L...5S}. A more detailed analysis based on Xtend will be presented in a forthcoming paper (Kurashima et al., in preparation).

\section{Comparison with theoretical calculations} \label{sec:discussion}

In this study, we for the first time demonstrated that the Fe-peak elemental composition within Cas A differs among the three Fe-rich blobs using the high-energy resolutional observations with XRISM/Resolve. In particular, the SE region exhibits high Mn/Cr and Ni/Fe ratios, whereas those ratios are significantly lower in the NW region, suggesting that explosion asymmetries may have generated these abundance differences. In this section, we compare our findings with a multidimensional supernova simulation to discuss which physical processes produce the asymmetric distribution of Fe-peak elements in Cas A.

Figure \ref{fig:f3} compares the elemental mass ratios derived from each region we observed with those in individual fluid elements of a two-dimensional SN model. This model is based on the explosion calculations of \cite{2011ApJ...738...61F} for a 21 $M_\odot$ progenitor calculated by \cite{2002ApJ...576..323R}, with an explosion energy of $\approx10^{51}~\mathrm{erg}$ \citep[see also][]{2021MNRAS.502.2319F}. The simulation implements a neutrino‐driven mechanism aided by standing accretion shock instability (SASI) and convection, thereby incorporating the asymmetric effects expected in core‐collapse SNe. The calculation was carried out up to $\sim1~$s after the explosion, and only those fluid elements that escape the gravitational potential of the proto-neutron star are included in the comparison.  Neutrino reactions are also included in the nucleosynthesis calculations, allowing their influence to be compared directly with the observational results. In deriving the elemental mass ratios, the calculations are carried out in the stable state after all decay processes of unstable isotopes have been completed.
For the comparison, we use the inner ejecta by selecting zones with an O/Fe mass ratio below 0.1. The small marker points in Figure \ref{fig:f3} represent the ejecta from the incomplete Si burning (i.e., quasi-equilibrium: QSE) and thus the large marker points reflect the composition of the complete Si burning region. The color of each model data point in the figure indicates the electron fraction, illustrating the transition from neutron-rich ejecta (light blue) to proton-rich ejecta (pink).  

Figure \ref{fig:f3} top left shows a comparison of the Ca/Fe and Mn/Cr ratios with theoretical predictions. Because Ca is produced primarily in the incomplete Si-burning region \citep[e.g.,][]{1985ApJ...295..604T,1995ApJS..101..181W}, its abundance serves as an indicator of the degree of contamination from that region. In particular, for XRISM/Resolve observations, the telescope's point spread function and the $0.5^{\prime}\times0.5^{\prime}$ size of the Resolve pixels \citep{2025PASJ..tmp...28T} make it difficult to isolate regions purely rich in Fe, and explosion-driven mixing is expected to blend compositions from multiple burning layers. 

The most observationally significant and interesting result is that the Mn abundance differs among the regions. In our current comparison, the Mn/Cr ratio in the SE region aligns with those in the incomplete Si-burning zone, whereas in the NW region it appears to match the ratios in the $\alpha$-rich freezeout (complete Si-burning) zone. This is exactly opposite to the trend indicated by the Ca/Fe ratio, where the higher Ca abundance in the NW region implies greater contamination from the incomplete Si-burning zone. Therefore, these observational results cannot be explained by mixing effects alone. If the Mn/Cr ratio reflects the nucleosynthesis in the $\alpha$-rich freezeout region, then this variation in Mn abundance may signal differences in the extent of neutrino processing. \cite{2023ApJ...954..112S} have also detected Mn in the Fe-rich structure in Cas A's southeast region and demonstrated that neutrino emission of (3--5)$\times10^{53}~\mathrm{erg}$ from a SN assuming a typical neutron star mass can account for its synthesis \citep[see also][]{2008ApJ...672.1043Y}. In the theoretical models we compared, the neutrino luminosity is of order of $10^{52}~\mathrm{erg}$, but if in reality it were an order of magnitude higher and spatially dependent, it might explain the asymmetric Mn/Cr distribution observed in Cas A. However, once mixing and overlap from other regions are taken into account, the current data set does not allow us to conclusively identify the underlying cause.

Figure \ref{fig:f3} top middle shows the distribution of another important neutron-rich element, Ni. In the case of Ni, different atomic codes (AtomDB and SPEX) currently yield different absolute abundances \citep[see][for detail]{2025PASJ...77S.171P}, making a discussion of its absolute value difficult. However, the trend that the Ni/Fe ratio is lowest in the NW region and highest in the SW region is consistent across both atomic codes. For example, the AtomDB-based evaluation finds a Ni/Fe mass ratio of $\sim$0.07--0.1, whereas SPEX shows values in the range of $\sim$0.03--0.06. Since almost all Ni is synthesized in the complete Si-burning region \citep[e.g.,][]{1985ApJ...295..604T,1995ApJS..101..181W}, this variation in Ni/Fe likely reflects differences in the physical conditions near the explosion center. In particular, the electron fraction, $Y_{\rm e}$ (i.e., neutron excess), is an important parameter for determining Ni/Fe, and $Y_{\rm e}$ is expected to vary within the SN interior due to two main factors: (a) differences in local neutrino irradiation and (b) variations in the progenitor's internal $Y_{e}$. 

Regarding the scenario (a), the 21 $M_{\odot}$ model we compared already incorporates these effects, its theoretical predictions span a wide range of $M_{\rm Ni}/M_{\rm Fe} \sim$ 0.03--1.0. For example, materials subjected to intense neutrino irradiation undergo charge-current interactions (i.e., $n + \nu_{e} \rightarrow p + e^{-}$), becoming proton-rich (high $Y_{e}$ of $>0.5$) \citep[e.g.,][]{2006ApJ...637..415F,2006PhRvL..96n2502F,2012ARNPS..62..407J,2018ApJ...852...40W,2018MNRAS.477.3091V,2024ApJ...964L..16B}, whereas materials in the innermost regions that have escaped from neutrinos could remain neutron-rich (low $Y_{e}$ of $<0.5$) in our calculations. The values derived from AtomDB are located at around the model’s peak region, while those from SPEX concentrate toward the lower end of this range. Observational values showing Ni/Fe mass ratios below 0.07 are found in both SPEX and AtomDB analyses, suggesting ejecta with $Y_{e}\approx0.5$ or higher. Explaining these observations may require either low initial metallicity of the progenitor as poposed in \cite{2020ApJ...893...49S} or proton-rich ejecta that have been heavily processed by neutrinos \citep{2021Natur.592..537S}. Interestingly, the observed values (especially those in the SE region) seem to agree well with the proton-rich ejecta predicted by the model. This may suggest that a large fraction of the innermost ejecta has been exposed to neutrinos strongly enough to become proton-rich. The bottom panels of Figure \ref{fig:f3} show a comparison with nucleosynthesis calculations in which the electron fractions were fixed at $Y_{e}= 0.51$, resembling a state where most of the ejecta become proton-rich as a result of neutrino irradiation. Interestingly, the mass ratios of Mn/Cr, Ni/Fe, Ti/Fe, and Cr/Fe are in good agreement with these proton-rich cases \citep[see also][]{2021Natur.592..537S}. Recent multi-dimensional simulations have further demonstrated that, especially in the case of massive progenitors, ejecta with $Y_{e} \geq 0.5$ can dominate \citep[e.g.,][]{2024ApJ...962...71W}, making the similarity to Cas A particularly noteworthy.

Regarding the scenario (b), multidimensional simulations of a massive star in the final evolutionary stages predict that intense oxygen burning drives significant mixing beyond convective boundaries and alters the internal distribution of the electron fraction, $Y_{\rm e}$ \cite[e.g.,][]{1994ApJ...433L..41B}. This phenomenon, known as ``shell merger'', has been reproduced in various theoretical calculations \citep[e.g.,][]{2011ApJ...733...78A,2019ApJ...881...16Y,2020ApJ...890...94Y,2021MNRAS.506L..20Y,2024MNRAS.533..687R}, and it has been suggested that Cas A’s progenitor also experienced such an inhomogeneous shell-merger mixing just before the SN explosion \citep{2025ApJ...984..185S,2025ApJ...990..103S}. Therefore, it would be plausible that this stellar mixing generated an asymmetric $Y_{\rm e}$ distribution in the progenitor, giving rise to the asymmetric Ni/Fe and Mn/Cr ratios.  Furthermore, this effect may have enhanced the explosion asymmetry of the supernova \citep[e.g.,][]{2017MNRAS.472..491M,2021ApJ...915...28B}.

The stable Ti we observed will also provide important insights into discussions of asymmetric explosions (Figure \ref{fig:f3} right). As with the well-known radioactive isotope $^{44}\mathrm{Ti}$ \citep[e.g.,][]{2014Natur.506..339G,2015Sci...348..670B}, stable Ti (predominantly $^{48}\mathrm{Ti}$) is synthesized in amounts that vary with the entropy at the explosion center, making it a valuable tracer of asymmetric explosions \citep[e.g.,][]{2021Natur.592..537S,2025ApJ...986...94S}. Recently, \cite{2021Natur.592..537S} reported the detection of stable Ti in the southeast Fe-rich region and argued that this structure represents high-entropy plumes produced by neutrino heating during the explosion. Despite the much lower photon statistics compared to previous Chandra observations, we leveraged XRISM/Resolve's superior energy resolution to reassess these synthesized abundances and obtained results that agree with the previous findings (see the SE values in Figure \ref{fig:f3} right). Although we did not achieve significant detections in the NW and SW regions at present, future deeper observations should enable us to discuss the distribution of Ti as well.

We note that the model employed in this work is a 2D (axisymmetric) simulation that was evolved up to $\sim$1 s after core bounce. In such 2D models, hydrodynamic instabilities such as neutrino-driven convection and SASI are artificially constrained by axial symmetry, producing coherent, axis-aligned plumes that differ from the more stochastic, multi-directional structures found in 3D simulations \citep[e.g.,][]{2012ApJ...755..138H,2014ApJ...786...83T,2016PASA...33...48M}. Moreover, since the model represents an early post-bounce phase, it does not include the later evolution involving the Ni-bubble effect \citep[e.g.,][]{2015A&A...577A..48W}, reverse shocks, and chemical mixing over hundreds of years \citep[e.g.,][]{2016ApJ...822...22O,2021A&A...645A..66O}. Therefore, our comparison with this model should be regarded as qualitative, aiming to illustrate possible trends in $Y_{\rm e}$-dependent nucleosynthesis rather than a one-to-one match to present-day ejecta morphology. We also note that recent three-dimensional simulations that follow the evolution from pre-collapse convection through core collapse and explosion \citep[e.g.,][]{2017MNRAS.472..491M,2021ApJ...915...28B,2025ApJ...982....9V} generally predict less coherent and more stochastic mixing, along with highly complex distributions of Fe-peak elements. In these models, multi-dimensional perturbations present in the Si/O shells before collapse are amplified during the explosion and lead to strongly asymmetric, small-scale structures in the Fe-rich ejecta, in stark contrast to the artificially coherent, axis-aligned plumes produced in 2D simulations.

\section{Summary and Conclusion} \label{sec:summary}

In this work, we have employed high-resolution X-ray spectroscopy with XRISM/Resolve to perform a detailed compositional analysis of Fe-peak elements (Ti, Cr, Mn, Ni, Fe) in three distinct Fe-rich regions (SE, NW, SW) of Cas A. Our observations reveal a significant detection ($>4\sigma$) of Mn He$\alpha$ emission in the SE region, with negligible Mn in the NW, alongside pronounced regional variations in Mn/Cr, Ni/Fe, and Ti/Fe ratios. It would be difficult to explain such asymmetries based solely on simple matter mixing and instead point to a combination of effects: (1) differential mixing of material from various Si-burning layers at the time of explosion, (2) spatial inhomogeneities in the electron fraction ($Y_{\rm e}$) within the progenitor and near the explosion center, and (3) locally dependent enhancements of $\nu$-process and $\nu p$-process nucleosynthesis driven by neutrino irradiation. These results provide new insights from a different perspective into previous studies of the asymmetry in Cas A \citep[e.g.,][]{2014Natur.506..339G,2015Sci...347..526M,2021Natur.592..537S}, suggesting that the asymmetric structures may have possessed different physical conditions during the explosion.

Comparison with multidimensional, neutrino-driven explosion models indicates that the observed abundance asymmetries could be linked to variations in the electron fraction ($Y_e$) and/or neutrino irradiation in the innermost ejecta. The high Mn/Cr ratio in the SE region is consistent with proton-rich conditions ($Y_e > 0.5$) produced by intense neutrino processing, whereas the suppressed Mn/Cr ratio in the NW region suggests $\alpha$-rich freeze-out. The observed diversity in Ni/Fe ratios also supports local $Y_e$ variations, either originating from asymmetric neutrino irradiation or from asymmetric mixing in the progenitor’s late evolutionary stages. In addition, the marginal detection of stable Ti in the SE region may support the high-entropy ejecta driven by neutrino heating as previously reported. These results highlight that both neutrino interactions and progenitor mixing processes likely played critical roles in producing the asymmetric distribution of Fe-peak elements in Cas A. 

Our findings offer crucial constraints on the asymmetric mechanisms at play in core-collapse supernovae and on the role of neutrino irradiation in shaping the earliest phases of element synthesis. Future efforts, combining finer spatial-resolution spectroscopy with XRISM’s expanded mapping capabilities, next-generation microcalorimeter missions \citep[e.g.,][]{2024SPIE13093E..27K}, and three-dimensional explosion simulations \citep[e.g.,][]{2024arXiv240812462O}, will be essential for quantitatively disentangling the physical processes driving asymmetry in supernova explosions.

\vspace{0.5cm}
\noindent 
{\bf Acknowledgments}: We thank all of the scientists, engineers, and technicians that built the XRISM satellite, operate the mission, and developed the software that we used to analyze the data in this paper. We also thank the anonymous referee for their careful reading of the manuscript and for constructive comments and suggestions that significantly improved the clarity and quality of this paper. This work was partly supported by Japan Society for the Promotion of Science Grants-in-Aid for Scientific Research (KAKENHI) Grant Numbers, JP23K25907 (AB), JP23K13128 (TS). This work was supported by the JSPS Core-to-Core Program (grant number: JPJSCCA20220002) (YT) and the Japan Society for the Promotion of Science Grants-in-Aid for Scientific Research (KAKENHI) Grant Number JP20K04009 (YT). PP acknowledges support from NASA XRISM grants 80NSSC18K0988 and 80NSSC23K1656, and the Smithsonian Institution and the Chandra X-ray Center through NASA contract NAS8-03060. MA and JV acknowledge financial support from NWO under grant number 184.034.002. This work was supported by the Mitsubishi Foundation.





\bibliography{sample631}{}
\bibliographystyle{aasjournal}



\end{document}